\documentclass[twocolumn,showpacs,showkeys,preprintnumbers,amsmath,amssymb]{revtex4}

\usepackage{graphicx}
\usepackage{dcolumn}
\usepackage{bm}

\begin{document}

\preprint{APS/123-QED}

\title{Charm and longitudinal structure functions with the KLN
model}

\author{F. Carvalho$^1,3$, F.O. Dur\~aes$^2$, F.S. Navarra$^3$ and S.
Szpigel$^2$ \\} \affiliation{$^1$Dep. de Metem\'atica e
Computa\c{c}\~ao - Faculdade de Tecnologia da Universidade do
Estado do Rio de Janeiro\\ Rodovia Presidente Dutra, km 298, Polo
Industrial, Resende, RJ, CEP 27.537-000
\\
$^2$Centro de Ci\^encias e Humanidades - Universidade
Presbiteriana Mackenzie, CEP 01302-907, S\~{a}o Paulo,
SP, Brazil\\
$^3$Instituto de F\'{\i}sica - Universidade de S\~{a}o Paulo, CEP
05315-970 S\~{a}o Paulo, SP, Brazil} \email{babi@if.usp.br //
duraes@mackenzie.br // navarra@if.usp.br // szpigel@mackenzie.br}



\date{\today}

\begin{abstract}
We use the Kharzeev-Levin-Nardi (KLN) model of the low $x$ gluon
distributions to fit recent HERA data on $F_L$ and $F^c_2 \,
(F^b_2)$. Having checked that this model gives a good description
of the data, we use it to predict $F_L$ and $F^c_2$ to be measured
in a future electron-ion collider. The results are similar to
those obtained with the de Florian-Sassot and
Eskola-Paukkunen-Salgado nuclear gluon distributions. The
conclusion of this exercise is that the KLN model, simple as it
is, may still be used as an auxiliary tool to make estimates both
for heavy ion and electron-ion collisions.
\end{abstract}

\pacs{PACS: 12.38.-t  24.85.+p  25.30.-c}

\keywords{Quantum Chromodynamics, Nuclear Gluon Distribution,
Shadowing Effect}

\maketitle

The small-$x$ regime of QCD has been intensely investigated in
recent years (for recent reviews see, e.g. \cite{cgc,gllm}). The
main prediction is a transition from the linear regime described
by the DGLAP dynamics to a non-linear regime where parton
recombination becomes important in the parton cascade and the
evolution is governed by a non-linear  equation. At very small
values of $x$ we expect to observe the saturation of the growth of
the gluon densities in hadrons and nuclei. One of the main topics
of hadron physics to be explored  in the new accelerators, such as
the LHC and possibly the future electron-ion collider is the
existence of this new component of the hadron wave function,
denoted Color Glass Condensate (CGC)  \cite{cgc}.

The search for signatures of the CGC has been subject of an active
research (for recent reviews see, e.g. \cite{cgc}). Saturation
models \cite{gllm,GBW,bgbk,kowtea,iim,fs} can successfully
describe HERA data in the  small $x$ and low $Q^2$ region.
Moreover, some  properties which appear naturally in the formalism
of the color glass condensate have been observed experimentally.
These include, for example, geometric scaling
\cite{bl,scaling,iimc} and the supression of high $p_T$ hadron
yields at forward rapidities in $dAu$ collisions
\cite{brahms,jamal,kkt,dhj,gkmn,buw}. However, it has been shown
that both geometric scaling \cite{caola} and high $p_T$ supression
\cite{kahana,vogt04} can be understood with other explanations,
not based on saturation physics.

In view of these (and others) results we may conclude that there
is some evidence for saturation at HERA and RHIC. However, more
definite conclusions are not yet possible. In order to
discriminate between these different models and test the CGC
physics, it would be very important to consider an alternative
search. To this purpose, the future electron-nucleus colliders
offers a promising opportunity
\cite{raju_ea1,raju_ea2,raju_ea4,raju_ea5,kgn}.

The color glass condensate is important in itself as a new state
of matter. However, apart from that, we need to know very well its
properties since the CGC forms the initial state of the fluid
created in nucleus-nucleus collisions. Any detailed simulation of
a heavy ion collision needs a realistic Ansatz for the initial
conditions. This would correspond to knowing accurately  the
unintegrated gluon distribution in the projectile and in the
target. These distributions will presumably be known in the
future, with the help of the results of deep inelastic scattering
(DIS) off nuclei. Even before experiment, one can try to calculate
the gluon density in the initial state of heavy ion collisions by
numerically solving the classical Yang-Mills equations, as done in
\cite{kras}. This is however very time consuming. Meanwhile, in
practical applications we need to use use models for these
distributions. Since these models are used as input for heavy
numerical calculations, they have to be simple.

One simple approach to saturation physics was developed by
Kharzeev, Levin and Nardi (KLN) in a series of papers \cite{kln},
where a simple model for the unintegrated gluon distribution was
proposed. It was used in many phenomenological applications. In
particular it was very successful when applied to hydrodynamical
simulations \cite{hirano04}, \cite{hirano042} and  \cite{adil05}.
In  \cite{hirano04} it was shown that hydrodynamical simulations
with the KLN model initial conditions were able to the describe
centrality, rapidity and energy dependence of charged hadron
multiplicities very well. Moreover these simulations could
reproduce the transverse momentum spectra of charged pions and
also the centrality dependence of the nuclear modification
factors.

In view of the success of the KLN model as an input for numerical
simulations, we think that it would be interesting to confront  it
with recent DIS data and, if necessary, change it in order  to get
a better agreement with these data. Of course, the KLN will remain
a phenomenological model to be replaced by something more
fundamental in the future. However, in a refined and still simple
version, it may be a very useful tool, capturing the essential
physics of gluon saturation and parameterizing the presently
available experimental information.

In this paper we apply the KLN model to deep inelastic scattering,
the domain where parton distributions have to be tested. If it
fails badly in reproducing DIS data, it must be discarded. Since
the KLN model gives only an Ansatz for the gluon distribution but
says nothing about quarks, it is difficult to use it to make
predictions for the most well known DIS observables, such as the
structure functions $F_2$. We must then look for quantities which
are dominated by the gluon content of the proton. These are the
charm (bottom) and longitudinal structure functions, $F_2^c\,
(F_2^b)$ and $F_L$ respectively.

Let us first discuss charm production and its contribution to the
structure function. In the last years, both the H1 and ZEUS
collaborations have measured the charm component $F_2^c$ of the
structure function at small $x$ and have found it to be a large
(approximately $25\%$) fraction of the total \cite{f2cdata}. This
is in sharp contrast to what is found at large $x$, where
typically $F_2^c/F_2\, \approx {\cal{O}}(10^{-2})$. This behavior
is directly related to the growth of the gluon distribution at
small-$x$.

In order to estimate the charm contribution to the structure
function we consider the formalism developed in \cite{grvc} where
the charm quark is treated as a heavy quark and its contribution
is given by fixed-order perturbation theory. This involves the
computation of the boson-gluon fusion process $\gamma^*g
\rightarrow c\overline{c}$. A $c\overline{c}$ pair can be created
by boson-gluon fusion when the squared invariant mass of the
hadronic final state is $W^2 \ge 4m_c^2$. Since $W^2 =
\frac{Q^2(1-x)}{x} + M_N^2$, where $M_N$ is the nucleon mass, the
charm production can occur well below the $Q^2$ threshold, $Q^2
\approx  4m_c^2$, at small $x$. The charm contribution to the
proton/nucleus structure function, in leading order (LO), is given
by \cite{grv95}
\begin{equation}
\frac{1}{x} F_2^c(x,Q^2,m_c^2) = e_c^2
\frac{\alpha_s(\mu^{2})}{2\pi} \int_{ax}^1 \frac{dy}{y}\,
C_{g,2}^c(\frac{x}{y},\xi)\,g(y,\mu^{2})
 \,\,,
\label{f2c}
\end{equation}
where $a=1+\xi$ ($\xi \equiv \frac{m_c^2}{Q^2}$) and the
renormalization scale $\mu$ is assumed to be either
$\mu^{2}=4m_c^2$   or $\mu^{2}=4m_c^2 + Q^2$. $C_{g,2}^c$ is the
coefficient function given by
\begin{eqnarray}
&&C_{g,2}^c(z, \xi) = \{ [z^2 + (1-z)^2 + 4\, \xi \,
z\,(1-3z)- 8\, \xi ^2 \,z^2] \nonumber\\
&\times& ln \,H + \beta\,[-1 + 8\, z\,(1-z) -4\, \xi \,
z\,(1-z)]\}\,\,,
\end{eqnarray}
where $\beta= 1 - \frac{4\,\xi \, z}{(1-z)}$ and $H =
\frac{1+\beta}{1-\beta}$.

The dominant uncertainty in QCD calculations comes from the
uncertainty in the charm quark mass. In this paper we assume $m_c
= 1.2\,$ GeV. In (\ref{f2c}) $g(y,\mu^2)$ is the gluon
distribution, which is usually taken from the CTEQ \cite{cteq},
MRST \cite{mrst} or GRV \cite{grv98} parameterizations. In what
follows we shall use the KLN Ansatz:
\begin{eqnarray}
\label{kln} xg(x,Q^2)= \left\{ \begin{array}{ll}
\frac{\kappa_0}{\alpha_s(Q_s^2)}
\, S \, Q^2 \,(1-x)^D & , \,\,\, \textrm{$Q^2<Q_s^2$} \,\,,\\
                                       \frac{\kappa_0}{\alpha_s(Q_s^2)}
\, S \, Q_s^2 \,(1-x)^D & , \,\,\, \textrm{$Q^2>Q_s^2$} \,\,.
                \end{array}\right.
\end{eqnarray}

In the above expression $S$ is the area of the target and
$\alpha_s$ is the running coupling $\alpha_s = {12}/[25 \pi \,
ln(Q^2/\Lambda^2)]$ (with $\Lambda=0.224$ GeV). In ({\ref{kln})
$D=4$ and $\kappa_0$ is a constant parameter to be adjusted by
requiring that the distribution $xg(x,Q^2)$ satisfies the momentum
sum rule $\int_0^1 dx \, x g(x,Q^2) = p$, where $p$ is the value
obtained with the GRV98 gluon density. $Q_s$ is the saturation
scale given by:
\begin{eqnarray}
\label{eqk:14} Q_s^2(x)=Q_0^2\left(\frac{x_0}{x}\right)^{\lambda}
\end{eqnarray}
where $Q_0^2=0.34~~ GeV^2$, $x_0=3.0 \times 10^{-3}$ and
$\lambda=0.25$.

In Fig. \ref{fig:fig1} we show $F_2^c$ as a function of $x$
obtained with the above expression and compared to the ZEUS and H1
data. Solid and dashed lines correspond to different choices for
the renormalization scale. We can observe that there is a
surprisingly good  agreement between the KLN model and the data,
especially considering that only minor changes in the parameters
were made, with respect to those found previously in the analysis
of RHIC data \cite{kln}. We can also obtain a reasonable
description of large $Q^2$ data, which is also surprising because
the KLN model has no DGLAP evolution, being tuned to the low $x$
and low $Q^2$ region of the phase space, where gluon saturation is
expected to occur. In Fig. \ref{fig:fig2} we show the results for
$F_2^b$ and compare them with the H1 data \cite{f2cbdata}. With
the exception of the points with $Q^2 = 200$ GeV$^2$, the
agreement with data is similar to the one found for $F_2^c$.

\begin{figure*}
\includegraphics{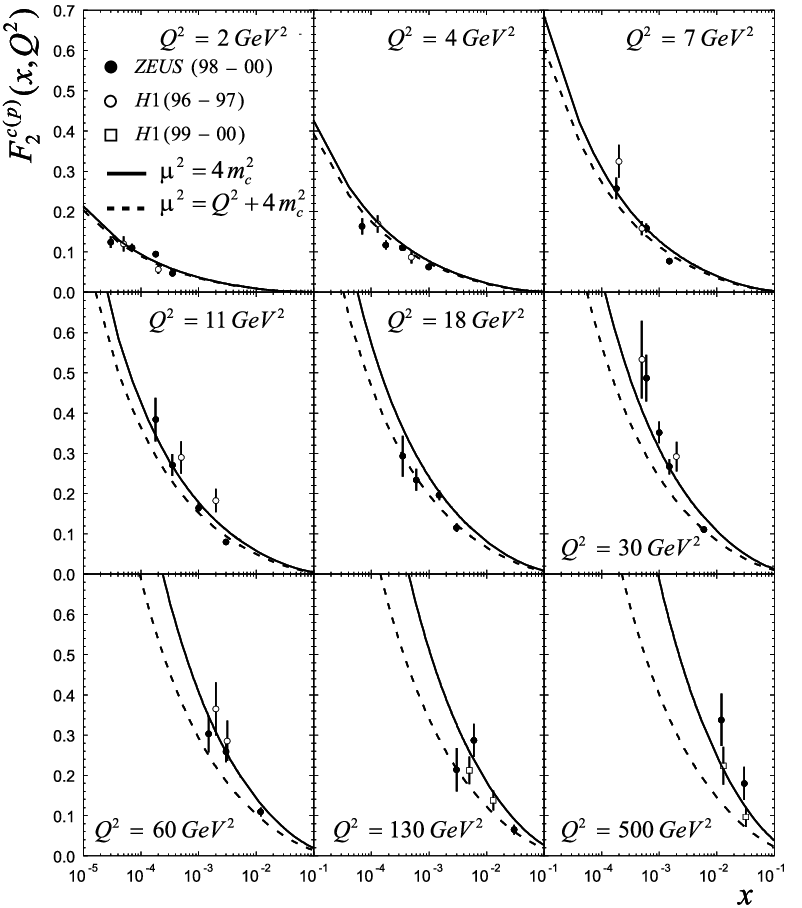}
\caption{\label{fig:fig1}Charm structure function, $F_2^c$
computed with the KLN model. Data are from Ref.
\protect\cite{f2cdata}.}
\end{figure*}

\begin{figure*}
\includegraphics{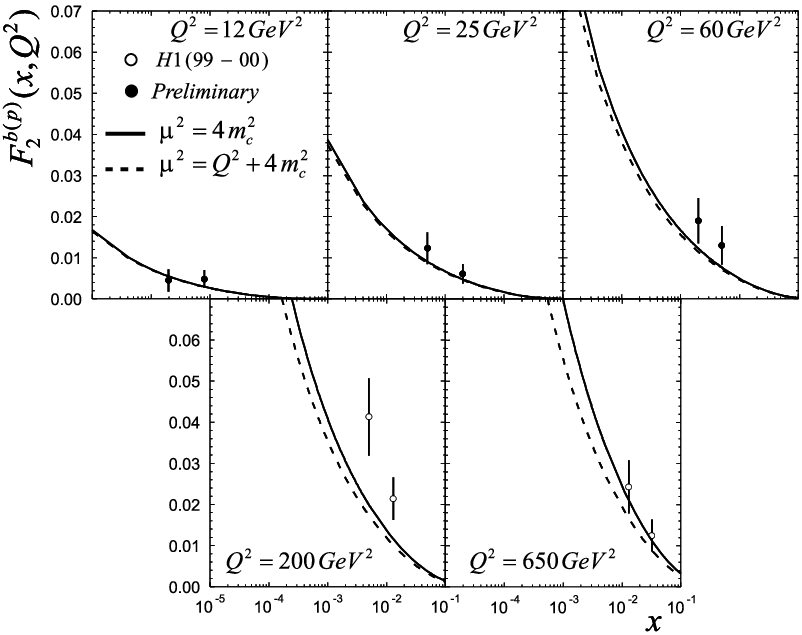}
\caption{\label{fig:fig2}Bottom structure function, $F_2^b$
computed with the KLN model. Data are from Ref.
\protect\cite{f2cdata}.}
\end{figure*}

New experimental HERA data on $F_L$ have recently appeared
\cite{fldata08}. The longitudinal structure function in deep
inelastic scattering is one of the observables from which the
gluon distribution can be unfolded. Longitudinal photons have zero
helicity and can exist only virtually. In the quark model,
helicity conservation of the electromagnetic vertex yields the
Callan-Gross relation, $F_L=0$, for scattering on quarks with spin
$1/2$ \cite{vicmagfl}. This does not hold when the quarks acquire
transverse momenta from QCD radiation.

Instead, QCD yields the Altarelli-Martinelli equation \cite{alta}
\begin{eqnarray}
F_L(x,Q^2) &=& \frac{\alpha_s(Q^2)}{2\pi}\,x^2 \int_x^1
\frac{dy}{y^3}\, [\frac{8}{3}\,F_2(y,Q^2) \nonumber\\
&+& 4\,\sum_q e_q^2 \,(1-\frac{x}{y})\,y\,g(y,Q^2)]\,\,,
\label{flalta}
\end{eqnarray}
expliciting the dependence of $F_L$ on the strong coupling
constant and the gluon density. At small $x$ the second term with
the gluon distribution is the dominant one. In Ref. \cite{cooper}
the authors have suggested that expression (\ref{flalta}) can be
reasonably approximated by $F_L(x,Q^2) \approx 0.3\, \frac{4
\alpha_s}{3 \pi} xg(2.5x,Q^2)$, which demonstrates the close
relation between the longitudinal structure function and the gluon
distribution.

In what follows we calculate $F_L$ using the Altarelli-Martinelli
equation, neglecting the $F_2$ contribution and using (\ref{kln})
in (\ref{flalta}). The results are presented in Fig.
\ref{fig:fig3}, where they are compared with the very recent H1
data \cite{fldata08} and with the results obtained with the help
of other standard gluon distribution functions. The agreement
between the KLN predictions and data is remarkably good. This is
again surprising because the KLN distribution is not expected to
work so well at such large values of $Q^2$. The good agreement
indicates that the KLN distribution has a good asymptotic behavior
and it is compatible both with the data and with the other
standard gluon distributions.

\begin{figure}
\includegraphics{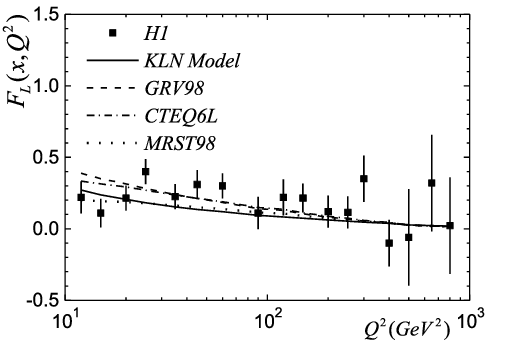}
\caption{\label{fig:fig3}Longitudinal structure function $F_L$
computed with the KLN model and with the gluon distributions taken
from CTEQ6L (dash dotted line), MRST98 (dotted line) and GRV98
(dashed line). Data are from Ref. \protect\cite{fldata08}}.
\end{figure}

\begin{figure}
\includegraphics{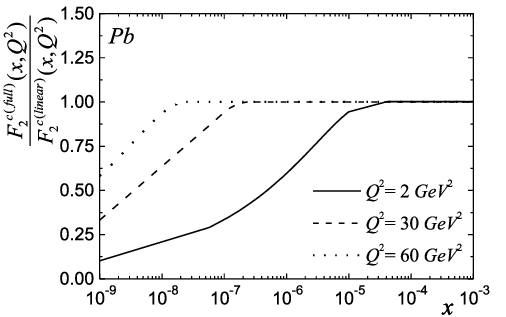}
\caption{\label{fig:fig4}Ratio full to linear (explained in the
text) $F^{c A}_{2}$ for $A=208$.}
\end{figure}

\begin{figure}
\includegraphics{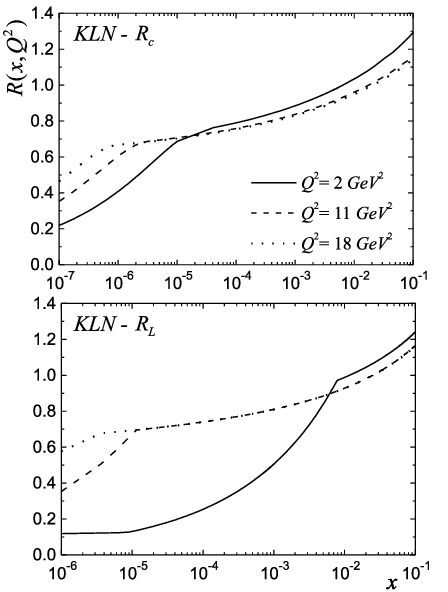}
\caption{\label{fig:fig5}Ratios $R_c$ (top) and $R_L$ (bottom)
predicted by the KLN model for $A=208$ and different values of
$Q^2$.}
\end{figure}

Having checked that the KLN distribution reproduces satisfactorily
the existing DIS data on electron-proton collisions, we shall now
use it to make predictions for electron-ion collisions. The
expression for the nuclear charm structure function $F^{c,A}_{2}$
is the same except for the change $g (y, Q^2) \rightarrow g^A (y,
Q^2)$, where $ g^A (y, Q^2)$ is obtained from Eq. (\ref{kln}) with
the replacements $S \rightarrow S^A = A^{2/3} S $ and $Q^2_s
\rightarrow Q^{2 A}_s$, where $ Q^{2 A}_s = A^{1/3} Q^2_s $. In
order to estimate the strength of non-linear effects in $eA$
processes, we can compute the linear contribution to $F^{c A}_{2}$
using only the second line of Eq. (\ref{kln}). We call this
$F_{2}^{c(linear)}$ and compare it with $F_{2}^{c(full)}$, where
the latter is calculated with both lines of Eq. (\ref{kln}) and
thus includes non-linear effects. The ratio $F_{2}^{c(full)} /
F_{2}^{c(linear)}$ as a function of $x$ is shown in Fig.
\ref{fig:fig4} for $A=208$ and for several values of $Q^2$.

The deviation of this ratio from unity shows the importance of
non-linear effects. As expected, for large $x$ and for large $Q^2$
there are no saturation effects. In fact, saturation effects are
only noticeable at $x < 10^{-6}$ and for small values of  $Q^2$.
This result confirms the findings of \cite{erike1}, where an
estimate of saturation effects in $eA$ collisions performed with
the color dipole approach also led to the conclusion that they are
only marginally visible in $F^{c,A}_{2}$. Having an idea of where
saturation effects could be relevant, we can compute an observable
quantity, which is $R_c(x,Q^2) = {F_{2 }^{c,A}(x,Q^2)}/ {A
F_{2}^{c,p}(x,Q^2)}$. The deviation from unity in this ratio is an
indication of saturation physics. A depletion in this ratio is
called ``shadowing'', whereas an enhancement is called
``anti-shadowing''.

In Fig. \ref{fig:fig5} on the top panel we calculate  $R_c$ and
estimate the magnitude of shadowing, which can be of 50 \% at very
low (but still reachable) values of $x$ and $Q^2$. The equivalent
ratio for the longitudinal structure functions $R_L(x,Q^2) =
{F_L^A(x,Q^2)} / {A F_L^p(x,Q^2)}$ is shown in the bottom panel of
Fig. \ref{fig:fig5} for the same choices of $Q^2$. We observe that
significant non-linear effects start to appear at larger values of
$x$. $R_L$ seems thus more promising than $R_c$. Following
\cite{erike1} we compare the normalized ratios $R_c$ and $R_L$
obtained with the KNL model with the same ratios computed with the
standard collinear factorization approach with nuclear parton
distribution functions (nPDF's). We take two extreme cases, one
with almost no shadowing at all, based on the nPDF's of de Florian
and Sassot (called here DS) \cite{ds04} and one with maximum
shadowing based on the Eskola, Paukkunen and Salgado nPDF's
(called here EPS) \cite{eps08}.

The ratios $R_c$ and $R_L$ are shown in Figs. \ref{fig:fig6} and
\ref{fig:fig7} respectively. From the figures we see that the KLN
model interpolates between the two extreme parameterizations, DS
and EPS, being closer to the latter. This is expected, since the
EPS gluon distribution comes from a fit of world data where BRAHMS
data on forward particle production were included. Both KLN and
EPS take RHIC data into account.

In summary, we have used the KLN model for the low $x$ gluon
distributions, slightly changing the parameters fixed from
previous analysis, to fit HERA data on $F_L$ and $F^c_2$. Having
checked that this model gives a good description of the data, we
have used it to predict $F_L$ and $F^c_2$ to be measured in
electron-ion collisions. The results are close to those obtained
with the DS and EPS nuclear gluon distributions. The conclusion of
this exercise is that the KLN model, simple as it is, may still be
used as an auxiliary tool to make estimates both for heavy ion and
electron-ion collisions.

\begin{figure}
\includegraphics{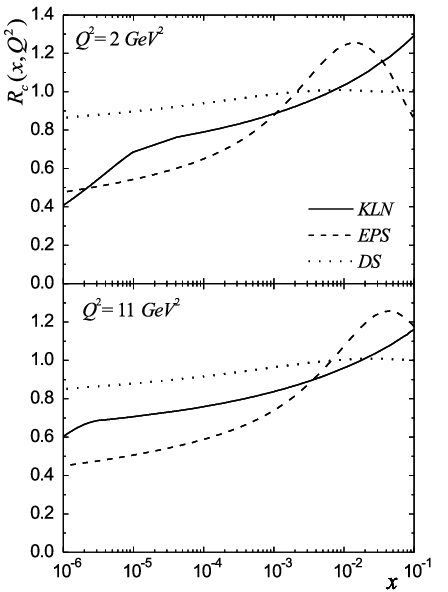}
\caption{\label{fig:fig6}Ratio $R_c$ calculated with the KLN, DS
and EPS nuclear gluon distribution functions for $A=208$ and
$Q^2=2$ (top) and $Q^2=11$ (bottom) GeV$^2$.}
\end{figure}

\begin{figure}
\includegraphics{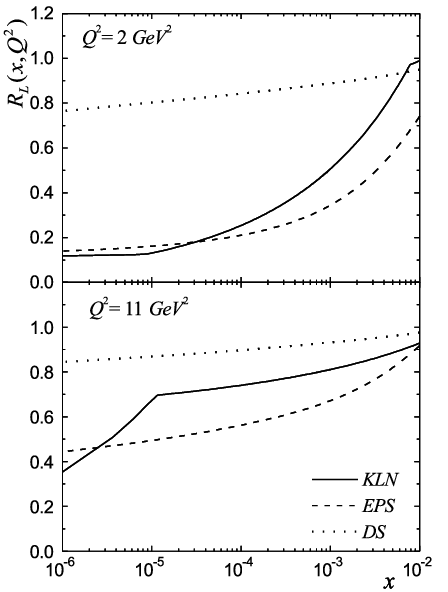}
\caption{\label{fig:fig7}Ratio $R_L = $ calculated with the KLN,
DS and EPS nuclear gluon distribution functions for $A=208$ and
$Q^2=2$ (top) and $Q^2=11$ (bottom) GeV$^2$.}
\end{figure}

\begin{acknowledgments}
This work was  partially financed by the Brazilian funding
agencies CNPq and FAPESP. S.S. and F.O.D. are grateful to the
Instituto Presbiteriano Mackenzie for the support through
MackPesquisa.
\end{acknowledgments}


\end{document}